\documentclass[12pt,a4paper]{article}
\usepackage{amsmath}
\usepackage{amssymb}
\usepackage{amsthm}
\usepackage{float}
\usepackage{amsfonts}
\usepackage{graphicx}
\usepackage{verbatim}
\usepackage[left=2cm,right=2cm,top=3cm,bottom=2.5cm]{geometry}
\usepackage[numbers]{natbib}
\usepackage[utf8]{inputenc}
\usepackage[usenames,dvipsnames,svgnames]{xcolor}
\usepackage[colorlinks=true,
      linkcolor=red,
      urlcolor=gray,
      citecolor=blue]{hyperref}

\def\myalign#1{%
  \def\trule{\noalign{\smallskip\hrule\medskip}}
  \def\nebc{\nearrow\bigcup}
  \def\sebc{\searrow\bigcup}
  \def\pminf{{}_{-\infty}|^{+\infty}}
  \let\Inf\infty
  \def\amp{&} 
  \vbox{\mathsurround0pt\openup1\jot
    \halign{%
      &$\displaystyle##\hfil\tabskip0pt$&\amp##\tabskip1em\crcr
      \noalign{\hrule height1pt\smallskip}#1\noalign{\smallskip\hrule height1pt}\crcr}}}
      
\begin{document}

\begin{center}
 \textbf{A study of perturbations in scalar–tensor theory using 1 + 3 covariant approach}
\end{center}
\hfill\newline
\begin{center}
Joseph Ntahompagaze$^{1,2,4}$, Amare Abebe$^{2,1,3}$ and Manasse R. Mbonye$^{4}$\\
Email: ntahompagazej@gmail.com,\; amare.abbebe@gmail.com,\; mmbonye@gmail.com\\
\end{center}
\hfill\newline
$^{1}$ Astronomy and Astrophysics Division, Entoto Observatory and Research Center, Ethiopia  \\
$^{2}$ Department of Physics, North-West University, South Africa\\
$^{3}$ Center for Space Research, North-West University, South Africa \\
$^{4}$ Department of Physics, College of Science and Technology, University of Rwanda, Rwanda

% \begin{history}
% \received{Day Month Year}
% \revised{Day Month Year}
% \end{history}

\begin{center}
 Abstract
\end{center}

This work discusses scalar-tensor theories of gravity, with a focus on the Brans-Dicke sub-class, and one that also takes note of the 
latter's equivalence with $f(R)$ gravitation theories. A $1 + 3$ covariant formalism is used in this case to discuss covariant perturbations
on a background Friedmann-Laima\^itre-Robertson-Walker (FLRW) space-time. Linear perturbation equations are developed, 
based on gauge-invariant gradient variables. Both scalar and harmonic decompositions are applied to obtain second-order equations. These equations can 
then be used for further analysis of the behavior of the perturbation quantities in such a scalar-tensor theory of gravitation. Energy density 
perturbations are studied for two systems, namely for a scalar fluid-radiation system and for a scalar fluid-dust system, for $R^{n}$ models.
For the matter-dominated era, it is shown that the dust energy density perturbations grow exponentially, a result which agrees 
with those already in existing literature. In the radiation-dominated era, it is found that the behavior of the radiation energy-density perturbations is oscillatory,
with growing amplitudes for $n > 1$, and with decaying amplitudes for $0 < n < 1$. This is a new result.\\
\hfill\newline
$keywords:$ $f(R)$ gravity -- scalar-tensor --scalar field --cosmology --covariant perturbation.\\
$PACS numbers:$ 04.50.Kd, 98.80.-k, 95.36.+x, 98.80.Cq;\\
$MSC numbers:$ 83Dxx, 83Fxx, 83Cxx

\section{Introduction}
Scalar-tensor theories have been widely explored \cite{singh1987general,fujii2003scalar,brans2008scalar,scalar2} and 
their relationship with $f(R)$ gravity theories are discussed \cite{7,scalar1,chakraborty2016solving}.
Recent focus \cite{scalar5,li2007cosmology,fara07,ntahompagaze2017f,sami2017reconstructing,sami2017inflationary} has been on the $f(R)$ theories as a special class of Brans-Dicke scalar-tensor theories.
In the current work we continue to explore this relationship following on our recent work \cite{ntahompagaze2017f}. 
In particular, we apply a $1+3$ covariant formalism to study cosmological perturbations. 
This formalism was introduced by Ellis and Bruni in the 1980s \cite{treciokas1971isotropic,ellis1989covariant}. This approach is constructed in 
such a way that the spacetime four-dimensional manifold is split into a 3-dimensional sub-manifold perpendicular to a time-like vector field.
This formalism has advantage that the quantities defined have both geometrical and physical meaning and they can be measured. 
In \cite{bruni1992cosmological}, the authors showed the importance of gauge-invariant gradient variables and their physical meaning.
In cosmology, perturbations of several quantities like the energy density parameter, expansion parameter and those of curvature have 
previously been treated. The $1+3$ covariant formalism has been applied to study linear perturbation in General 
Relativity \cite{ellis1989covariant}.
The same studies has been done in $f(R)$ gravity at linear order and results have been obtained \cite{amare2,27}.
In scalar-tensor theory, the $1+3$ covariant linear perturbation has been studied \cite{SanteCarloni4, osano2007gravitational}, but it was
limited to vacuum case. The assumption that the hyper-surfaces are with constant scalar field was made in that study. 
Those hyper-surfaces are perpendicular to a vector field.

In this paper, we consider hyper-surfaces with constant curvature. They have been considered for $f(R)$
covariant perturbations \cite{27,amare2,amare4}. This consideration is taken due to the equivalence
between metric $f(R)$ theory of gravity and Brans-Dicke scalar-tensor theory. 
The equivalence between $f(R)$ gravity and scalar-tensor theory in five 
dimensions have been considered in \cite{chakraborty2016solving,chakraborty2016gravity}, where Jordan frame was given much attention and bulk consideration which 
resulted in hyper-surfaces of four dimensional space-times. In our study, the hyper-surfaces are constructed from $1+3$ covariant decomposition
of spacetime such that they have three dimensions.
The work presented in this paper is a follow-up of the work previously done by the authors \cite{ntahompagaze2017f}, where the equivalence between $f(R)$ theory and
scalar-tensor theory has been explored. Here, this equivalence is extended to covariant linear perturbations for two fluid system
with the consideration that scalar field behaves like a fluid (scalar fluid). The two fluid system (radiation-scalar field or dust-scalar field) 
are considered with the motivation that towards the end of a scalar field driven inflation the universe experienced a mixture of
scalar field and radiation. Later during the matter dominated epoch one can think of a situation where the scalar field was so small that 
its effect becomes less significant.

This paper is organized as follows: in the next section, the $1+3$ covariant formalism is described. In Section \ref{fRST}, the
equivalence between $f(R)$ and scalar tensor theories of gravitation is reviewed. In Section \ref{BACKGROUND}, background evolution 
equations for the FLRW universe are provided to play a ground for the coming perturbation equations. Section \ref{GRADIENT} is for definition
of gradient variables to be used in evolution equations. The next two sections are for linear and scalar evolution equations.
In Section \ref{HARMONIC}, we perform harmonic decomposition to make equations simpler where partial differential equations are reduced
ordinary differential equations. In Section \ref{scalardominance1}, we apply perturbations to scalar field dominated universe 
and two different cases have been considered; namely
scalar field-dust system and scalar field-radiation system. Finally, the conclusion is drawn in Section \ref{Conclusion1}.
The adopted spacetime signature is $(-+++)$ and unless stated otherwise, 
we have used the convention $8\pi G=c=1$, where $G$ is the gravitational constant and $c$ is the speed of light.

\section{1+3 covariant approach for scalar-tensor theories}\label{covaiantapproach}
The $1+3$ covariant approach divides space-time into foliated hyper-surfaces and a perpendicular 4-vector-field.
In this process, the cosmological manifold $(\mathcal{M},g)$ is decomposed into the sub-manifold $(M,h)$ which has a 
perpendicular 4-velocity field vector $u^{a}$ \cite{treciokas1971isotropic,ellis1989covariant}. 
The background under consideration is the FLRW spacetime with a
scalar field $\phi$. Note that these hypersurfaces are spacelike. The 4-velocity field vector $u^{a}$ is defined as
\begin{equation}
u^{a}=\frac{dx^{a}}{d \tau}\; ,
\end{equation} 
where $\tau$ is proper time such that $u_{a}u^{a}=-1.$
This approach helps in the decomposition of the metric $g_{ab}$ into the projection tensor $h_{ab}$ as \cite{amare4,3}:
\begin{equation}
g_{ab}= h_{ab}-u_{a}u_{b}\; .
\end{equation} 
Here the projection tensors (parallel $U^{a}_{b}$ and orthogonal $h_{ab}$) are defined as 
\begin{equation}
U^{a}_{b}=-u^{a}u_{b} \Rightarrow U^{a}_{c}U^{c}_{b}=U^{a}_{b},\; U^{a}_{a}=1,\; U_{ab}u^{b}=u_{a}\; ,
\end{equation}
and 
\begin{equation}
h_{ab}=g_{ab}+u_{a}u_{b} \Rightarrow h^{a}_{c}h^{c}_{b}=h^{a}_{b},\; h^{a}_{a}=3,\; h_{ab}u^{b}=0\; .
\end{equation}
The projection tensor $U^{a}_{b}$ projects parallel to the 4-velocity vector $u^{a}$ and the $h_{ab}$ is responsible 
for the metric properties of instantaneous rest-spaces of observers moving perpendicularly  with 4-velocity $u^{a}$.
The derivatives also follow with respect to those projectors. The covariant time derivative (for a given tensor $T^{ab}_{cd}$) is given as
\begin{equation}
\dot{T}^{ab}_{cd}=u^{e}\nabla_{e}T^{ab}_{cd}\; .
\end{equation} 
In this approach, the kinematic quantities which are obtained from irreducible parts of the decomposed $\nabla_{a}u_{b}$ are given as \cite{3}
\begin{equation}
\nabla_{a}u_{b}=\tilde{\nabla}_{a}u_{b}-u_{a}\dot{u}_{b}=\frac{1}{3}\Theta h_{ab}+\sigma_{ab}+\omega_{ab}-u_{a}\dot{u}_{b}\; , \label{definitionofu}
\end{equation}
where $\tilde{\nabla}$ refers to a 3-spatial gradient, the volume rate of expansion of the fluid $\Theta =\tilde{\nabla}_{a}u^{a}$, the Hubble parameter $H=\frac{\Theta}{3}=\frac{\dot{a}}{a}$, 
the rate of distortion of the matter flow $\sigma_{ab}=\tilde{\nabla}_{<a}u_{b>}$, is the trace-free symmetric rate of the shear 
tensor $(\sigma_{ab}=\sigma_{(ab)},\sigma_{ab}u^{b}=0, \sigma^{a}_{a}=0)$.
The vorticity tensor $\omega_{ab}=\tilde{\nabla}_{[a}u_{b]}$ is the skew-symmetric vorticity tensor and describes the rotation of the matter 
relative to a non-rotating frame.
The relativistic acceleration vector (not that of the expansion of the universe) is given as
$\dot{u}^{a}=u^{b}\nabla_{b}u^{a}$. These kinematic quantities provide informations about the overall spacetime kinematics. 
We define also the volume element for 3-restspace as 
\begin{equation}
\eta_{abc}=u^{d}\eta_{abcd}, \text{  } \eta_{abc}=\eta_{[abc]}, \text{  } \eta_{abc}u^{c}=0\; , 
\end{equation}
where $\eta_{abcd}$ is a 4-dimensional volume element.
The matter energy-momentum tensor $T_{ab}$ is also decomposed with the 1+3 covariant approach and it is given as \cite{amare2,3}
\begin{equation}
T_{ab}=\mu u_{a}u_{b}+q_{a}u_{b}+u_{a}q_{b}+ph_{ab}+\pi_{ab}\; ,\label{13}
\end{equation}
where $\mu$ is the relativistic energy density,
$q^{a}$ is the relativistic momentum density (energy flux relative to $u^{a}$),
$p$ is relativistic isotropic pressure, and
$\pi_{ab}$ is the trace-free anisotropic pressure of the fluid ($\pi^{a}_{a}=0, \pi_{ab}=\pi_{(ab)}, \pi_{ab}u^{b}=0$).
The energy-momentum tensor for the perfect fluid can be recovered by setting ($q^{a}=\pi_{ab}=0$) and that tensor $T_{ab}$ reads
\begin{equation}
T_{ab}=\mu u_{a}u_{b}+ph_{ab}\; ,
\end{equation} 
where the equation of state for perfect fluid is $p=p(\mu)$.
\section{$f(R)$ theory in scalar-tensor language}\label{fRST}
Let us consider the action that represents $f(R)$ gravity given as
\begin{equation}
I=\frac{1}{2\kappa}\int d^{4}x\sqrt{-g}\left[f(R)+2\mathcal{L}_{m}\right]\; ,
\end{equation}
where $\kappa= 8\pi G$,$R$ is Ricci scalar and $\mathcal{L}_{m}$ is the matter Lagrangian.
The action in scalar-tensor theory has the form \cite{scalar5,7}
\begin{equation}
I_{f(\phi)}=\frac{1}{2\kappa}\int d^{4}\sqrt{-g}\left[f(\phi)+2\mathcal{L}_{m}\right]\; ,\label{frstt1}
\end{equation}
where $f(\phi)$ is the function of $\phi(R)$ and specifically where we define 
the scalar field $\phi$ to be \cite{scalar5,ntahompagaze2017f}
of the form
\begin{equation}
\phi=f'-1\; ,\label{phifRandST}
\end{equation}
where for GR, we have a vanishing scalar field.
Here the prime indicates differentiation with respect to $R$ and we require that the 
scalar field $\phi$ be invertible \cite{scalar1,7,scalar3}.
The field equations from the action in Eq. \eqref{frstt1} are given in \cite{7} as:
\begin{equation}
G_{ab}=\frac{\kappa}{\phi+1}T^{m}_{ab}+\frac{1}{(\phi+1)}\left[\frac{1}{2} g_{ab}\Big(f-(\phi+1)R\Big)
+\nabla_{a}\nabla_{b}\phi-g_{ab}\square \phi\right]\;,\label{frstt2a}
\end{equation}
where $\square =\nabla_{c}\nabla^{c}$ is the covariant D'Alembert operator and $T^{m}_{ab}$ is the matter energy-momentum tensor.
The scalar field $\phi$ obeys the Klein-Gordon equation \cite{scalar5}
\begin{equation}
\square \phi -\frac{1}{3}\Big(2f-(\phi+1)R+T^{m}\Big)=0 \; , \label{KG}
\end{equation}
where $T^{m}$ is the trace of the matter energy momentum tensor. \\
One can consider at this stage the Friedmann and Rychaudhuri equations given as
\begin{equation}
\Theta^{2}=3(\overline{\mu}_{m}+\mu_{\phi})-\frac{9K}{a^{2}}\; ,\label{fried}
\end{equation}
and 
\begin{equation}
\dot{\Theta}+\frac{1}{3}\Theta^{2}+\frac{1+3w}{2}\overline{\mu}_{m}+\frac{1}{2}(\mu_{\phi}+3p_{\phi})=0\; ,\label{duri}
\end{equation}
 where $w$ is equation of state parameter (EoS), $\mu_{\phi}$ is the energy density and $p_{\phi}$ is isotropic pressure of
scalar fluid respectively; $K$ stands for curvature and has the values $0,-1,+1$ for flat, open and closed universe respectively,
$a$ is the scale factor and $\overline{\mu}_{m}=\frac{\mu_{m}}{\phi+1}$. For the flat $(K=0)$ FLRW background one has Ricci scalar given as
\begin{equation}
 R=6\Big(\frac{\ddot{a}}{a}+\frac{\dot{a}^{2}}{a^{2}}\Big)\; .
\end{equation}

\section{Background (zeroth-order) quantities}\label{BACKGROUND}
Now since the background is considered to be the FLRW spacetime, the background quantities: energy density 
and isotropic pressure for the curvature fluid are given as \cite{amare4}
\begin{eqnarray}
&&\mu_{R}=\frac{1}{f'}\Big[\frac{1}{2}(Rf'-f)-\Theta f'' \dot{R}\Big]\\
&&p_{R}=\frac{1}{f'}\Big[\frac{1}{2}(f-Rf')+f''\ddot{R} +f'''\dot{R}^{2}+\frac{2}{3}\Theta f''\dot{R}\Big]\; .
\end{eqnarray}
Analogously, we have energy density and pressure for scalar fluid as 
\begin{eqnarray}
&&\mu_{\phi}=\frac{1}{\phi+1}\left[\frac{1}{2}\Big((\phi+1)R-f\Big)-\Theta\dot{\phi}\right]\; ,\label{14aa}\\
&&p_{\phi}= \frac{1}{\phi+1}\left[\frac{1}{2}\Big(f-R(\phi+1)\Big)+\ddot{\phi}-\frac{\dot{\phi}\dot{\phi}'}{\phi'}\phi''
+\frac{\phi''\dot{\phi}^{2}}{\phi'^{2}}+\frac{2}{3}\Theta \dot{\phi}\right]\; .\label{14bb}
\end{eqnarray}
For FLRW spacetime,  the kinematic quantities presented in Eq. \eqref{definitionofu} satisfy
\begin{equation}
 \sigma_{ab}=0, \text{  } \omega_{a}=0, \text{   } \dot{u}_{a}=0, \text{  } \tilde{\nabla}_{a}\Theta=0\; .
 \label{kinematicquantitybackg}
\end{equation}
Also for any scalar quantity $f$, in the background we have
\begin{equation}
 \tilde{\nabla}_{a}f=0,
\end{equation}
and hence, at the background
\begin{equation}
\tilde{\nabla}_{a}\mu=0, \text{   }
\tilde{\nabla}_{a}p=0, \text{  }
\tilde{\nabla}_{a}\phi=0\; , \label{nablapsi}
\end{equation}
and 
\begin{equation}
q^{\phi}_{a}=\pi^{\phi}_{ab}=0\; .
\end{equation}
The energy conservation or simply the continuity equations for matter and scalar field fluid are given as
\begin{equation}
\dot{\mu}_{m}=-\Theta(\mu_{m}+p_{m})\; , 
\end{equation}
\begin{equation}
\dot{\mu}_{\phi}=-\Theta(\mu_{\phi}+p_{\phi})+\frac{\dot{\phi}\mu_{m}}{(\phi+1)^{2}}\; . \label{yyayaya}
\end{equation}
One can easily notice the coupling between these two equations from matter energy density in the last term of equation \eqref{yyayaya} where $\mu_{m}$ is 
present as a factor.

\section{Definition of gradient variables}\label{GRADIENT}
Gauge-invariant quantities are key for the cosmological perturbations analysis. A quantity is said to be 
gauge-invariant (GI), (Stewart-Walker Lemma \cite{amare4,stewart1974perturbations}) 
if it vanishes in the background.
We define GIs in 1+3 covariant perturbations in the following way. The GI variable that characterizes energy density perturbation in spatial variation
is given as \cite{27,amare4,SanteCarloni1,bruni1992gauge}
\begin{equation}
X_{a}=\tilde{\nabla}_{a}\mu\; . 
\end{equation}
Then from this quantity, we can define the following GI variable
\begin{equation}\label{gradientDa}
D_{a}=\frac{a}{\mu}X_{a}=\frac{a}{\mu}\tilde{\nabla}_{a}\mu\; . 
\end{equation}
The ratio $\frac{a}{\mu}$ helps to evaluate the magnitude of energy density perturbations relative to the background.
Further, We define two other quantities. The spatial gradient of the volume expansion
\begin{equation}
Z_{a}=a\tilde{\nabla}_{a}\Theta\; , \label{Za} 
\end{equation}
and spatial gradient of the 3-Ricci scalar $\tilde{R}$ as
\begin{equation}\label{gradientCa}
C_{a}=a^{3}\tilde{\nabla}_{a}\tilde{R}\; .
\end{equation}
This 3-Ricci scalar $\tilde{R}$ is given as \cite{amare4}
\begin{equation}\label{tildeR}
 \tilde{R}=-\frac{2}{3}\Theta^{2} +\frac{2\mu_{m}}{\phi+1}+\mu_{\phi}\; .
\end{equation}
Finally, we define two other gradient variables $\Phi_{a}$ and $\Psi_{a}$ that characterize perturbation due to scalar field and momentum of
scalar field as 
\begin{equation}\label{gradientPhia}
 \Phi_{a}=a\tilde{\nabla}_{a}\phi\; ,
\end{equation}
\begin{equation}\label{gradientPsia}
\Psi_{a} =a\tilde{\nabla}_{a}\dot{\phi}\; .
\end{equation}
The above defined gradient variables can be used to derive $1+3$ covariant perturbations equations for the universe. 
In the following section, we provide linear evolution equations for these gradient variables.
\section{Linear evolution equations}\label{LIINEAR}
The linear temporal evolution equation for gradient variable responsible for
energy density perturbation is obtained from Eq. \eqref{gradientDa} and is given as \cite{27,amare4,SanteCarloni1}
\begin{equation}\label{firstdotDam}
\dot{D}^{m}_{a}=-(1+w)Z_{a}+w\Theta D^{m}_{a}\; . 
\end{equation}
The linear evolution equation for comoving volume expansion $Z_{a}$is obtained by taking time derivative of Eq. \eqref{Za}
and is given as
\begin{equation}\label{Zanew}
\begin{split}
\dot{Z}_{a}&=\Big(\frac{\dot{\phi}}{\phi+1}-\frac{2\Theta}{3}\Big)Z_{a}-\frac{w\tilde{\nabla}^{2}D^{m}_{a}}{(1+w)}
-\frac{\tilde{\nabla}^{2}\Phi_{a}}{\phi+1}+\Big[\frac{1}{2\phi'}+\frac{\mu_{m}}{(\phi+1)^{2}}-\frac{f}{2(\phi+1)^{2}}\\
&-\frac{\Theta\dot{\phi}}{(\phi+1)^{2}}-\frac{\phi''\tilde{\nabla}^{2}R}{\phi'(\phi+1)}
+\frac{\phi'\tilde{\nabla}^{2}R}{(\phi+1)^{2}}-\frac{2K}{(\phi+1)a^{2}}\Big]\Phi_{a}
+\Big[\frac{\phi w-1}{(w+1)(\phi+1)}\mu_{m}\\
&-\frac{w}{1+w}\Big(-\frac{1}{3}\Theta^{2}+\frac{f}{2(\phi+1)}+\frac{\Theta\dot{\phi}}{\phi+1}
-\frac{\phi'}{\phi+1}\tilde{\nabla}^{2}R\Big)\Big]D^{m}_{a}+\frac{\Theta}{\phi+1}\Psi_{a}\; .
\end{split}
\end{equation}
For the scalar field, we have the evolution equation of gradient variable $\Phi_{a}$ obtained from Eq. \eqref{gradientPhia} as
\begin{equation}
\dot{\Phi}_{a}=\Psi_{a}-\frac{w\dot{\phi}}{w+1}D^{m}_{a}\; .
\end{equation}
In this equation, one can easily see that the time derivative of $\Phi_{a}$ couples with gradient variables
which characterize energy density and scalar field, $D^{m}_{a}$ and $\Psi_{a}$ respectively. 
The evolution equation for the momentum of scalar field is obtained from Eq. \eqref{gradientPsia} as
\begin{equation}
\begin{split}
 \dot{\Psi}_{a}=&\frac{\ddot{\phi}'}{\phi'}\Phi_{a}-\frac{w\ddot{\phi}}{(w+1)}D^{m}_{a}\; .
\end{split}
 \end{equation}
The constraint equation that connects gradient variables obtained from Eqs. \eqref{tildeR} and \eqref{gradientCa} to be
\begin{equation}
\begin{split}
\frac{C_{a}}{a^{2}}&=-\Big(\frac{4}{3}\Theta +\frac{2\ddot{\phi}}{\phi+1}\Big)Z_{a}+\frac{2\mu_{m}}{\phi+1}D^{m}_{a}
-\Big(\frac{2\mu_{m}}{(\phi+1)^{2}}-\frac{f}{(\phi+1)^{2}}
+\frac{2\Theta}{(\phi+1)}\frac{\ddot{\phi}'}{\phi'}\\
&-2\frac{\Theta\ddot{\phi}}{(\phi+1)}-\frac{4K}{a^{2}(\phi+1)}
-2\frac{\phi''\tilde{\nabla}^{2}R}{(\phi+1)\phi'}+2\frac{\phi'\tilde{\nabla}^{2}R}{(\phi+1)^{2}}
\Big)\Phi_{a}+2\frac{1}{\phi+1}\tilde{\nabla}^{2}\Phi_{a}\; . \label{linearconstraintCa11single}
\end{split}
\end{equation}
This constraint equation shows how gradient variables $Z_{a}$,$D^{m}_{a},\Phi_{a}$ and $\Psi_{a}$ are related at linear level.
Equations $\eqref{Zanew}-\eqref{linearconstraintCa11single}$ are new, together with Eq. \eqref{firstdotDam}, they
describe the evolution of the gradient variables defined
in the previous section.

\section{Scalar equations}\label{SCALAR}
Scalar perturbations are believed to be the ones responsible for large-scale structure formation \cite{ellis1990density}. 
Here we extract the scalar part from the quantities under consideration.
Hence we provide the local decomposition for a quantity $X_{a}$ as \cite{27,amare4}
\begin{equation}
a\tilde{\nabla}_{b}X_{a}=X_{ab}=\frac{1}{3}h_{ab}X+\Sigma^{X}_{ab}+X_{[ab]}\; , 
\end{equation}
where $\Sigma^{X}_{ab}=X_{(ab)}-\frac{1}{3}h_{ab}X$ describes shear and $X_{[ab]}$ describes vorticity. 
Now let us apply the comoving differential operator $a\tilde{\nabla}_{a}$ to $D_{a},Z_{a}$ and $C_{a}$ to have
\begin{equation}\label{scalargradientvariabels1}
\Delta_{m}=a\tilde{\nabla}^{a}D^{m}_{a}, \text{  } Z=a\tilde{\nabla}^{a}Z_{a}, \text{ and  } C=a\tilde{\nabla}^{a}C_{a}\; ,
\Psi =a\tilde{\nabla}^{a}\Psi_{a}, \text{ and  } \Phi=a\tilde{\nabla}^{a}\Phi_{a}\; . 
\end{equation}
The scalar evolution equations of the gradient variables defined in Eq. \eqref{scalargradientvariabels1} are therefore given as
\begin{equation} 
\dot{\Delta}_{m}=-(1+w)Z+w\Theta \Delta_{m}\; , \label{dotDeltasingle1}
\end{equation}
\begin{equation}
\begin{split}
\dot{Z}&=\Big(\frac{\dot{\phi}}{\phi+1}-\frac{2\Theta}{3}\Big)Z-\frac{w\tilde{\nabla}^{2}\Delta_{m}}{(1+w)}
-\frac{\tilde{\nabla}^{2}\Phi}{\phi+1}+\Big[\frac{1}{2\phi'}+\frac{\mu_{m}}{(\phi+1)^{2}}-\frac{f}{2(\phi+1)^{2}}\\
&-\frac{\Theta\dot{\phi}}{(\phi+1)^{2}}-\frac{\phi''\tilde{\nabla}^{2}R}{\phi'(\phi+1)}
+\frac{\phi'\tilde{\nabla}^{2}R}{(\phi+1)^{2}}-\frac{2K}{(\phi+1)a^{2}}\Big]\Phi +\Big[\frac{\phi w-1}{(w+1)(\phi+1)}\mu_{m}\\
&-\frac{w}{1+w}\Big(-\frac{1}{3}\Theta^{2}+\frac{f}{2(\phi+1)}+\frac{\Theta\dot{\phi}}{\phi+1}
-\frac{\phi'}{\phi+1}\tilde{\nabla}^{2}R\Big)\Big]\Delta_{m}+\frac{\Theta}{\phi+1}\Psi \; ,\label{dotZsingle1}
\end{split}
\end{equation}
\begin{equation}
\dot{\Phi}=\Psi-\frac{w\dot{\phi}}{w+1}\Delta_{m}\; ,\label{dotPhisingle1}
\end{equation}
\begin{equation}
\begin{split}
 \dot{\Psi}&=\frac{\ddot{\phi}'}{\phi'}\Phi-\frac{w\ddot{\phi}}{(w+1)}\Delta_{m}\; .\label{dotPsisingle1}
\end{split}
\end{equation}
Eq. \eqref{dotDeltasingle1} can be obtained in the literature \cite{27,amare4,SanteCarloni1}.
The constraint equation is
\begin{equation}
\begin{split}
\frac{C}{a^{2}}&=-\Big(\frac{4}{3}\Theta +\frac{2\ddot{\phi}}{\phi+1}\Big)Z+2\frac{\mu_{m}}{\phi+1}\Delta_{m}
+\Big(\frac{f-2\mu_{m}}{(\phi+1)^{2}}
-\frac{2\Theta}{(\phi+1)}\frac{\ddot{\phi}'}{\phi'}\\
&+\frac{2\Theta\ddot{\phi}}{(\phi+1)}+\frac{4K}{a^{2}(\phi+1)}
+\frac{2\phi''\tilde{\nabla}^{2}R}{(\phi+1)\phi'}-\frac{2\phi'\tilde{\nabla}^{2}R}{(\phi+1)^{2}}
\Big)\Phi+2\frac{1}{\phi+1}\tilde{\nabla}^{2}\Phi\; .\label{Csingle1}
\end{split}
\end{equation}
From the above scalar equations, we obtained second-order equations in 
$\Delta_{m}$ and in $\Phi$ as
\begin{equation}
\begin{split}
&\ddot{\Delta}_{m}=\Big(\frac{\dot{\phi}}{\phi+1}-\frac{2\Theta}{3}+w\Theta\Big)\dot{\Delta}_{m}
+\Big[-w\Theta\Big(\frac{\dot{\phi}}{\phi+1}-\frac{2\Theta}{3}\Big)+\frac{\phi w-1}{(\phi+1)}\mu_{m}
-w\Big(1+\frac{1}{3}\Theta^{2}\\
&-\frac{f}{2(\phi+1)}+\frac{\phi'\tilde{\nabla}^{2}R}{\phi+1}-\dot{\Theta}\Big)\Big]\Delta_{m}
+w\tilde{\nabla}^{2}\Delta_{m}+\frac{(1+w)}{\phi+1}\tilde{\nabla}^{2}\Phi-(1+w)\Big[\frac{1}{2\phi'}
+\frac{\mu_{m}}{(\phi+1)^{2}}\\
&-\frac{f}{2(\phi+1)^{2}}-\frac{\Theta\dot{\phi}}{(\phi+1)^{2}}-\frac{\phi''\tilde{\nabla}^{2}R}{\phi'(\phi+1)}
+\frac{\phi'\tilde{\nabla}^{2}R}{(\phi+1)^{2}}-\frac{2K}{(\phi+1)a^{2}}\Big]\Phi
-\frac{(1+w)\Theta}{\phi+1}\dot{\Phi}\; ,
\end{split}
\end{equation}
\begin{equation}
\begin{split}
\ddot{\Phi}&=\frac{\ddot{\phi}'}{\phi'}\Phi-\frac{w\dot{\phi}}{w+1}\dot{\Delta}_{m}-\frac{2w\ddot{\phi}}{w+1}\Delta_{m}\; ,
\end{split}
\end{equation}
where $w$ is constant with time.
These two second-order differential equations will be analyzed later after obtaining their harmonic decomposed counterparts.
\section{Harmonic decomposition}\label{HARMONIC}
The harmonic decomposition approach is a way of obtaining eigenfunctions with the corresponding wavenumber 
for a harmonic oscillator differential equation after applying separation of variables to that second-order differential equation. 
The differential equation can be represented as \cite{amare2}:
\begin{equation}
\ddot{X}+A_{1}\dot{X}+A_{2}X=A_{3}(Y,\dot{Y})\; ,\label{harmonic1} 
\end{equation}
where $A_{1},A_{2}$ and $A_{3}$ are independent of $X$ (in this work, these coefficients are time dependent) and they represent damping, restoring and source terms respectively. The separation
of variables for solutions of Eq. \eqref{harmonic1} is done such that $X(\vec{x})$ and $Y(\vec{x})$ depend on spatial 
variable $\vec{x}$ only, and $X(t)$ and $Y(t)$ depend on time variable $t$ only, so that
\begin{equation}
X(\vec{x},t)=X(\vec{x})X(t), \text{   and  } Y(\vec{x},t)=Y(\vec{x})Y(t)\; . 
\end{equation}
To bring the eigenfunctions $Q_{k}$ in the game, we make a summation (or integration) over wavenumber $k$ as
\begin{equation}
X=\sum_{k}X^{k}(t)Q_{k}(\vec{x}),   \text{  and  } Y=\sum_{k}Y^{k}(t)Q_{k}(\vec{x})\; , 
\end{equation}
where $Q_{k}$ are the eigenfunctions of the covariant Laplace-Beltrami operator such that
\begin{equation}
 \tilde{\nabla}^{2}Q=-\frac{k^{2}}{a^{2}}Q\; ,
\end{equation}
and the order of harmonic (wavenumber) $k$ is 
\begin{equation}
 k=\frac{2\pi a}{\lambda}\; , \label{wavenumberk}
\end{equation}
where $\lambda$ is the physical wavelength of the mode. The eigenfunctions $Q$ are time independent. That means 
\begin{equation}
 \dot{Q}(x)=0\; .
\end{equation}
This method of harmonic decomposition has been extensively used for 1+3 covariant linear
perturbations \cite{amare2,SanteCarloni4,amare4}. This approach 
allows one to treat the scalar perturbation equations as ordinary differential equations at each mode $k$ separately. Therefore, the analysis
becomes easier when dealing with ordinary differential equations rather than the former (partial differential) equations.
With this approach one has Eqs. \eqref{dotDeltasingle1}-\eqref{dotPsisingle1} rewritten as 
\begin{equation}
\dot{\Delta}^{k}_{m}=-(1+w)Z^{k}+w\Theta \Delta^{k}_{m}\; , \label{harmonicdotDeltasingle1}
\end{equation}
\begin{equation}
\begin{split}
\dot{Z}^{k}&=\Big(\frac{\dot{\phi}}{\phi+1}-\frac{2\Theta}{3}\Big)Z^{k}
+\Big[\frac{1}{2\phi'}+\frac{\mu_{m}}{(\phi+1)^{2}}-\frac{f}{2(\phi+1)^{2}}-\frac{\Theta\dot{\phi}}{(\phi+1)^{2}}
+\frac{k^{2}\phi''R}{a^{2}\phi'(\phi+1)}\\
&-\frac{k^{2}\phi'R}{a^{2}(\phi+1)^{2}}-\frac{2K}{(\phi+1)a^{2}}+\frac{k^{2}}{a^{2}(\phi+1)}\Big]\Phi^{k}
+\Big[\frac{\phi w-1}{(w+1)(\phi+1)}\mu_{m}-\frac{w}{1+w}\Big(-\frac{\Theta^{2}}{3}\\
&+\frac{f}{2(\phi+1)}+\frac{\Theta\dot{\phi}}{\phi+1}
+\frac{k^{2}\phi'R}{a^{2}(\phi+1)}\Big)+\frac{k^{2}w}{a^{2}(1+w)}\Big]\Delta^{k}_{m}+\frac{\Theta}{\phi+1}\Psi^{k}\; ,
\label{harmonicdotZsingle1}
\end{split}
\end{equation}
\begin{equation}
\dot{\Phi}^{k}=\Psi^{k}-\frac{w\dot{\phi}}{w+1}\Delta^{k}_{m}\; ,\label{harmonicdotPhisingle1}
\end{equation}
\begin{equation}
\begin{split}
 \dot{\Psi}^{k}&=\frac{\ddot{\phi}'}{\phi'}\Phi^{k}-\frac{w\ddot{\phi}}{(w+1)}\Delta^{k}_{m}\; .\label{harmonicdotPsisingle1}
\end{split}
\end{equation}
Eq. \eqref{harmonicdotDeltasingle1} can be obtained in the literature \cite{27,amare4,SanteCarloni1}.
The second-order equations are given as
\begin{equation}
\begin{split}
&\ddot{\Delta}^{k}_{m}-\Big(\frac{\dot{\phi}}{\phi+1}-\frac{2\Theta}{3}+w\Theta\Big)\dot{\Delta}^{k}_{m}
-\Big[w\Theta\Big(\frac{2\Theta}{3}-\frac{\dot{\phi}}{\phi+1}\Big)+\frac{\phi w-1}{(\phi+1)}\mu_{m}
-w\Big(1+\frac{\Theta^{2}}{3}\\
&-\frac{f}{2(\phi+1)}
-\frac{k^{2}\phi'R}{a^{2}(\phi+1)}\Big)+w\dot{\Theta}-\frac{k^{2}w}{a^{2}}\Big]\Delta^{k}_{m}=
-(1+w)\Big[\frac{1}{2\phi'}+\frac{\mu_{m}}{(\phi+1)^{2}}-\frac{f}{2(\phi+1)^{2}}\\
&-\frac{\Theta\dot{\phi}}{(\phi+1)^{2}}+\frac{k^{2}\phi''R}{a^{2}\phi'(\phi+1)}
-\frac{k^{2}\phi'R}{a^{2}(\phi+1)^{2}}-\frac{2K}{(\phi+1)a^{2}}+\frac{k^{2}}{a^{2}(\phi+1)}\Big]\Phi^{k}
-\frac{(1+w)\Theta}{\phi+1}\dot{\Phi}^{k}\; ,\label{secondorderHarmonicDeltam1}
\end{split}
\end{equation}
\begin{equation}
\begin{split}
\ddot{\Phi}^{k}&=\frac{\ddot{\phi}'}{\phi'}\Phi^{k}-\frac{w\dot{\phi}}{w+1}\dot{\Delta}^{k}_{m}-\frac{2w\ddot{\phi}}{w+1}\Delta^{k}_{m}\; .
\end{split}\label{secondorderHaromicPsi1}
\end{equation}
The set of equations \eqref{secondorderHarmonicDeltam1}-\eqref{secondorderHaromicPsi1} 
are the ones responsible for the evolution of
linear perturbations of single content (and scalar fluid component) of the perturbative universe in scalar-tensor theory of gravitation. 
One has to be reminded that these
harmonic decomposed quantities are associated with their wavenumber (or modes) $k$
with corresponding wavelength $\lambda$. The two parameters $k$ and $\lambda$ are connected with cosmological scale factor $a$ 
via equation \eqref{wavenumberk}. Thus, we can in principle make our analysis simplified by considering different scales such
as sub-horizon (short wavelength limit) scale and so on. 
The other interest might come from studying the quasi-Newtonian approximations
of these evolution equations. In the following, we will be studying the possible way of specifying fluids which are more
suitable with different approximation approaches. The reader should be reminded that, for perturbations during inflation epoch, one has to 
apply slow-roll approximation during the analysis of the behavior of comoving quantities.

\section{Scalar field dominated universe}\label{scalardominance1}
We can think of an epoch in the evolution of the universe for which the scalar field was dominating over the dust-like content
in flat FLRW spacetime cosmology.
In such a condition the universe was dominated by scalar field content and hence the matter fluctuations (perturbations) can be studied 
during such an epoch. Furthermore, we have assumed that those two fluids are non-interacting. We can neglect the term $\Phi \approx 0$
and consequently $\Psi \approx 0$ together with $\dot{\Phi}\approx 0$ and $\dot{\Psi}\approx 0$, with the support that scalar field is part of the background, hence its perturbation has no much interest from 
the homogeneity point of view. 
With this assumption Eqs. \eqref{harmonicdotDeltasingle1}- \eqref{harmonicdotZsingle1} are reduced to 
\begin{eqnarray}
&&\dot{\Delta}^{k}_{m}=-(1+w)Z^{k}+w\Theta \Delta^{k}_{m}\; , \label{dotDeltatwofluid1}\\
&&
\begin{split}
\dot{Z}^{k}&=\Big(\frac{\dot{\phi}}{\phi+1}-\frac{2\Theta}{3}\Big)Z^{k}
+\Big[\frac{\phi w-1}{(w+1)(\phi+1)}\mu_{m}+\frac{w}{1+w}\Big(\frac{1}{3}\Theta^{2}-\frac{f}{2(\phi+1)}\\
&-\frac{\Theta\dot{\phi}}{\phi+1}-\frac{k^{2}\phi'R}{a^{2}(\phi+1)}\Big)+\frac{k^{2}w}{a^{2}(1+w)}\Big]\Delta^{k}_{m}
+\frac{\Theta}{\phi+1}\Psi^{k}\; ,\label{dotZtwofluid1}
\end{split}
\end{eqnarray}
We therefore have second-order differential equation in $\Delta^{k}_{m}$ as
\begin{equation}
 \begin{split}
 \ddot{\Delta}^{k}_{m}&= \Big(\frac{\dot{\phi}}{\phi+1}-\frac{2\Theta}{3} +w\Theta\Big)\dot{\Delta}^{k}_{m}
+\Big[\Big(\frac{2\Theta}{3}-\frac{\dot{\phi}}{\phi+1}\Big)w
-\frac{(\phi w-1)}{(\phi+1)}\mu_{m}\\
&+w\Big(-\frac{1}{3}\Theta^{2}+\frac{f}{2(\phi+1)}+\frac{\Theta\dot{\phi}}{\phi+1}+\frac{k^{2}\phi'R}{a^{2}(\phi+1)}\Big)
+\frac{\Theta}{\phi+1}\frac{w\dot{\phi}}{w+1}
+w\dot{\Theta}-\frac{k^{2}w}{a^{2}}\Big]\Delta^{k}_{m}\; .
\end{split}
\end{equation}
At this stage, we can consider $R^{n}$ models, where the scale factor $a$ admits an exact solution of the form \cite{26} 
\begin{equation}\label{scalefactor}
a=a_{0}\Big(\frac{t}{t_{0}}\Big)^{\frac{2n}{3(1+w)}}\; , 
\end{equation}
where $a_{0}$ is the scale factor at the time where scalar field and dust energy densities were in equal.
This is a solution of scale factor of a particular orbit of the cosmological dynamical system treated in \cite{26}. 
In most cases $a_{0}$ 
and $t_{0}$ are normalized to unity \cite{amare4}. For a given scale factor such as the one in Eq. \eqref{scalefactor}, one has
the volume expansion $\Theta$, Ricci-scalar $R$ and matter energy density $\mu_{m}$ given from field equations as
\begin{eqnarray}
&&\Theta=\frac{2n}{(1+w)t}\; ,\label{Thetaaa}\\
&&R=\frac{4n[4n-3(1+w)]}{3(1+w)t^{2}}\; ,\label{Raaa} \\
&&\mu_{m}=\Big(\frac{3}{4}\Big)^{1-n}\Big[\frac{4n-3n(1+w)}{(1+w)^{2}t^{2}}\Big]^{n-1}\frac{4n^{3}-2n(n-1)[2n(3w+5)
-3(1+w)]}{3(1+w)^{2}t^{2}}\; . \label{muu}
\end{eqnarray}
By considering the above quantities, we have assumed that the $f(R)=\beta R^{n}$ is the one of $R^{n}$ models with normalized 
coefficient $\beta=1$. Note that these solutions have been obtained from dynamical system analysis of $R^{n}$ models \cite{26}.

\subsection{Scalar field-dust system}
\begin{figure}[h!]
\centering
\includegraphics[scale=0.6]{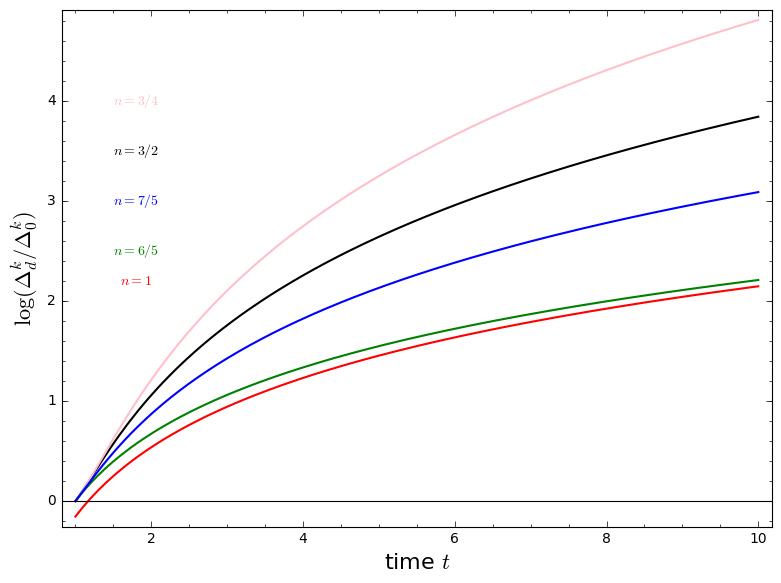}
\caption{Solution to Eq. \eqref{finaDust} for different values of $n$ with assumption that $\Delta^{k}_{(d)0}=10^{-3},\dot{\Delta}^{k}_{(d)0}=10^{-3}$}
\label{Figdifferentndust2}
\end{figure}

If we assume that the matter content in the system is dust-like, $w=0$, then we have perturbation equations as
\begin{equation}
\dot{\Delta}_{d}=-Z\; . \label{dotDeltatwofluid1dust}
\end{equation}
The scalar linear evolution equation for $Z$ is given by
\begin{equation}
\begin{split}
\dot{Z}&=\Big(\frac{\dot{\phi}}{\phi+1}-\frac{2\Theta}{3}\Big)Z-\frac{\mu_{d}}{(\phi+1)}\Delta_{d}\; .
\label{dotZtwofluid1dust}
\end{split}
\end{equation}
The second-order equation becomes
\begin{equation}
\begin{split}
 \ddot{\Delta}^{k}_{d}-\Big(\frac{\dot{\phi}}{\phi+1}-\frac{2\Theta}{3}\Big)\dot{\Delta}_{d}-\frac{\mu_{d}}{\phi+1}\Delta_{d}=0\; . \label{finaDust1}
\end{split}
\end{equation}
For $R^{n}$ models, the scale factor evolves as 
\begin{equation}
a=t^{\frac{2n}{3}}\; , 
\end{equation}
in dust-like matter content. With this assumption Eqs \eqref{Thetaaa}- \eqref{muu} reduce to 
\begin{equation}
\Theta=\frac{2n}{t}\; ,\label{Thetaredust1}
\end{equation}
\begin{equation}
R=\frac{4n(4n-3)}{3t^{2}}\; , 
\end{equation}
\begin{equation}
\mu_{d}=\Big(\frac{3}{4}\Big)^{1-n}\Big[\frac{4n^{2}-3n}{t^{2}}\Big]^{n-1}\frac{(16n^{3}+26n^{2}-6n)}{3t^{2}}\; .
\end{equation}
The scalar field has the form
\begin{equation}
\phi=n\left(\frac{4n(4n-3)}{3t^{2}} \right)^{n-1}-1\; .
\end{equation}
Therefore, we update Eq. \eqref{finaDust1} as
\begin{equation}
\begin{split}
 \ddot{\Delta}^{k}_{d}+\frac{10n-6}{3t}\dot{\Delta}_{d}
 -\frac{1}{3^{n}t^{2}}(-16n^{2}+26n-6)\Delta_{d}=0\; . \label{finaDust}
\end{split}
\end{equation}
In the following, we are going to explore some solutions for different values of $n$ considered in the literature, see for example 
Ref. \cite{3,27}.
If $n=1$ that is GR case, this equation reduces to 
\begin{equation}
\begin{split}
 \ddot{\Delta}^{k}_{d}+\frac{4}{3t}\dot{\Delta}^{k}_{d}-\frac{4}{3t^{2}}\Delta^{k}_{d}=0\; . \label{finaDustGRcase}
\end{split}
\end{equation}
This second-order differential equation admits a solution of the form
\begin{equation}
\Delta^{k}_{d}(t)=C_{1}t +C_{2}t^{-\frac{4}{3}}\; . \label{solutionfortwofluidGR}
\end{equation}
By assuming that 
at $\Delta^{k}_{d}(t=t_{0}=1)=\Delta^{k}_{0}$ and $\dot{\Delta}^{k}_{d}(t=t_{0}=1)=\dot{\Delta}^{k}_{0}$, 
we have constants of integration as
\begin{equation}
\Delta^{k}_{(d)0}=C_{1}+ C_{2}\; , \label{equationforDeltaeq}
\end{equation}
and 
\begin{equation}
\dot{\Delta}^{k}_{(d)0}=C_{1}-\frac{4}{3}C_{2}\; . \label{equationfordotDeltaeq}
\end{equation}
Thus solving the Eqs. \eqref{equationforDeltaeq} and \eqref{equationfordotDeltaeq} simultaneously, the constants are related by the following
two equations
\begin{equation}
C_{1}=\frac{4}{7}\Delta^{k}_{(d)0}+\frac{3}{7}\dot{\Delta}^{k}_{(d)0}\; ,
\end{equation}
and 
\begin{equation}
C_{2}=\frac{3}{7}(\Delta^{k}_{(d)0}-\dot{\Delta}^{k}_{(d)0})\; . 
\end{equation}
For the case where $n\neq 1$, we consider five cases. The choice of values of
$n$ was made following how it is made in \cite{amare4}, that is $n=3/2, n=4/3, n=3/4, n=6/5, n=7/5$. The solutions that correspond to these values
of $n$ are plotted in Fig. $\ref{Figdifferentndust2}$.
The solution presented in Eq. \eqref{solutionfortwofluidGR} 
is plotted in Fig. $\ref{Figdifferentndust2}$ together with other solutions for different value of 
$n$. Note that for the GR case, we do not have a scalar field $\phi=f'-1=0$ since $f'=1$. 
This is the consequence of Eq. \eqref{phifRandST}. For $n=3/4$, the $R^{n}$ dust models have a vanishing Ricci-scalar solution.
The behavior of solutions presented in Fig. $\ref{Figdifferentndust2}$ is
the same as those already exists in the literatures, see for example \cite{amare4,3}.

\subsection{Scalar field-radiation system}
For a universe composed of a mixture of the scalar field together with the radiation matter content, we can still assume 
that the background is filled with a scalar field in a flat FLRW spacetime. Since for radiation $w_{r}=\frac{1}{3}$,
we have the perturbation equations as
\begin{equation}
\dot{\Delta}^{k}_{r}=-\frac{4}{3}Z^{k}+\frac{\Theta}{3} \Delta^{k}_{r}\; , \label{dotDeltatwofluid1r}
\end{equation}
\begin{equation}
\begin{split}
\dot{Z}^{k}&=\Big(\frac{\dot{\phi}}{\phi+1}-\frac{2\Theta}{3}\Big)Z^{k}
+\Big[\frac{(\phi-3)}{4(\phi+1)}\mu_{r}-\frac{1}{4}\Big(-\frac{1}{3}\Theta^{2}+\frac{f}{2(\phi+1)}\\
&+\frac{\Theta\dot{\phi}}{\phi+1}+\frac{k^{2}\phi'R}{a^{2}(\phi+1)}\Big)+\frac{k^{2}}{4a^{2}}\Big]\Delta^{k}_{r}
+\frac{\Theta}{\phi+1}\Psi^{k}\; ,\label{dotZtwofluid1r}
\end{split}
\end{equation}
\begin{equation}
\begin{split}
 \dot{\Psi}^{k}&=-\frac{\ddot{\phi}}{4}\Delta^{k}_{r}\; .\label{dotPsitwofluid1r}
\end{split}
\end{equation}
We therefore have equation in $\Delta^{k}_{r}$ as
\begin{equation}
\begin{split}
 &\ddot{\Delta}^{k}_{r}- \Big(\frac{\dot{\phi}}{\phi+1}-\frac{\Theta}{3}\Big)\dot{\Delta}^{k}_{r}
-\Big[\frac{2\Theta}{9}-\frac{\dot{\phi}}{3(\phi+1)}
-\frac{(\phi-3)\mu_{r}}{3(\phi+1)}-\frac{1}{9}\Theta^{2}+\frac{f}{6(\phi+1)}\\
&+\frac{7\Theta\dot{\phi}}{12(\phi+1)}+\frac{k^{2}\phi'R}{3a^{2}(\phi+1)}
+\frac{\dot{\Theta}}{3}-\frac{k^{2}}{3a^{2}}\Big]\Delta^{k}_{r}=0\; .\label{twofluidDeltaradiation1}
\end{split}
\end{equation}
At this stage, if we consider $R^{n}$ models, where we can take the exact single fluid background transient solution as
\begin{equation}
a=t^{\frac{n}{2}}\; . 
\end{equation}
In this context, Eqs \eqref{Thetaaa}- \eqref{muu} reduce to
\begin{equation}
\Theta=\frac{3n}{2t}\; , 
\end{equation}
\begin{equation}
R=\frac{3n(n-1)}{t^{2}}\; ,
\end{equation}
\begin{equation}
\mu_{r}=\Big(\frac{3}{4}\Big)^{2-n}\Big(\frac{9n(n-1)}{4t^{2}}\Big)^{n-1}\Big(\frac{-5n^{3}+8n^{2}-2n}{t^{2}}\Big)\; ,
\end{equation}
and scalar field has the form
\begin{equation}
\phi =n\left(\frac{3n(n-1)}{t^{2}}\right)^{n-1}-1\; .
\end{equation}
We can write Eq. \eqref{twofluidDeltaradiation1} as
\begin{equation}
\begin{split}
 &\ddot{\Delta}^{k}_{r}+A_{1}(t)\dot{\Delta}^{k}_{r}+A_{2}(t)\Delta^{k}_{r}=0\; ,\label{twofluidDeltaradiation21}
\end{split}
\end{equation}
where 
\begin{equation}
A_{1}(t)=\frac{(-5n+4)}{2t}\; , 
\end{equation}
and 
\begin{equation}
\begin{split}
&A_{2}(t)= -\frac{(-3n^{2}+3n-2)}{3t}-\frac{(-11n^{2}+3n-4)}{4t^{2}}-\frac{k^{2}(n-2)}{3t^{n}}\\
&+\frac{3^{1-n}}{16t^{2}}\Big[n\Big(\frac{4n(4n-4)}{16/3}\Big)^{n-1}t^{-2(n-1)}-2\Big](-20n^{3}+32n^{2}-8n)\; .
\end{split}
\end{equation}

\begin{figure}[h!]
\centering
\includegraphics[scale=0.55]{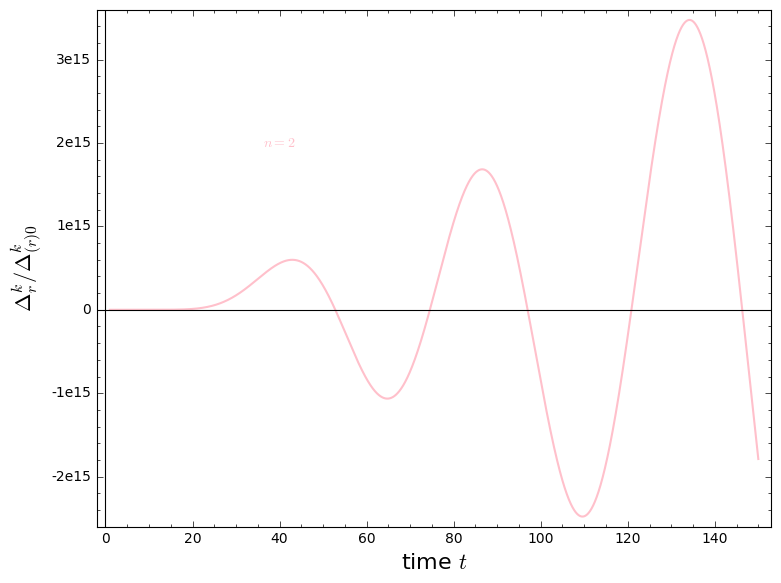}
\caption{Solution of fractional radiation energy density fluctuations for $n=2$, with assumption that
$\Delta^{k}_{(d)0}=10^{-4},\dot{\Delta}^{k}_{(d)0}=10^{-4}$}
\label{Figtwofluidradiation2}
\end{figure}

\begin{figure}[h!]
\centering
\includegraphics[scale=0.55]{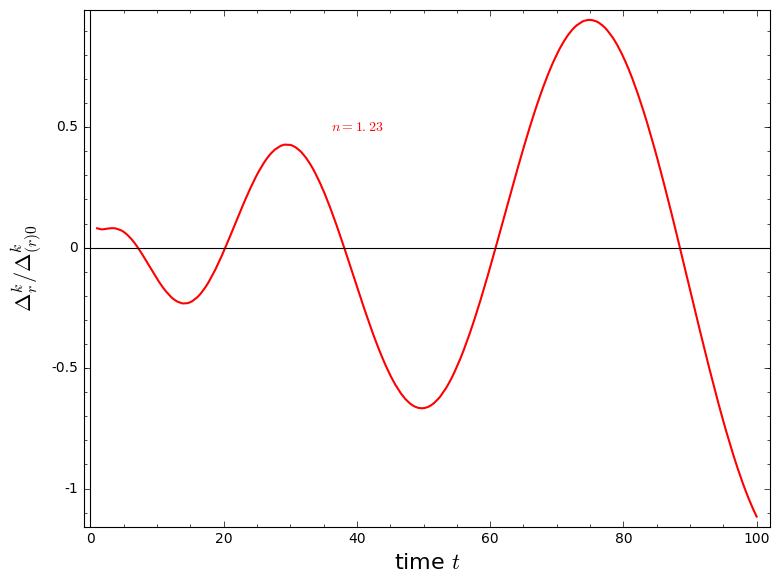}
\caption{Solution of fractional radiation energy density fluctuations for $n=1.23$, with assumption that
$\Delta^{k}_{(d)0}=10^{-4},\dot{\Delta}^{k}_{(d)0}=10^{-4}$}
\label{Figtwofluidradiation123}
%\end{figure}

%\begin{figure}[h!]
\centering
\includegraphics[scale=0.55]{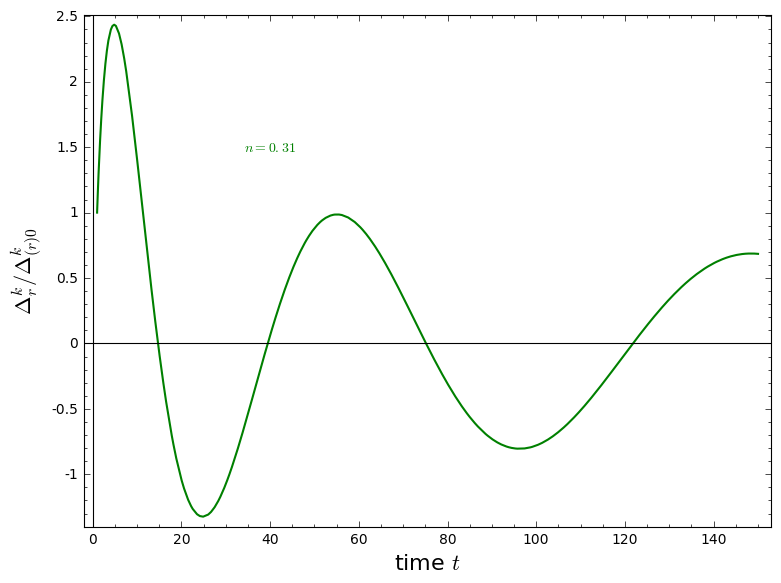}
\caption{Solution of fractional radiation energy density fluctuations for $n=0.31$, with assumption that
$\Delta^{k}_{(d)0}=10^{-4},\dot{\Delta}^{k}_{(d)0}=10^{-4}$}
\label{Figtwofluidradiation031}
\end{figure}

The numerical solutions of Eq. \eqref{twofluidDeltaradiation21} are obtained for $n=2$, $n=1.23$ and $n=0.31$ and plotted in Fig.\ref{Figtwofluidradiation2},
 Fig. \ref{Figtwofluidradiation123} 
and Fig.  \ref{Figtwofluidradiation031} respectively. 
During the numerical computation, the assumption of long wavelength is made for simplicity; that is, we have assumed that
$k^{2}$ is small enough compared to other terms (or $\frac{k^{2}}{a^{2}H^{2}}<<1$). However, the behavior of this solution is oscillatory
with changing in amplitude. Since we are dealing with the system where the scalar field was dominating in flat FLRW spacetime,
the treatment was that
the radiation energy perturbations are the ones expected to rise. Thus, surprisingly, it turns out that these perturbations grow in an oscillatory
shape. The oscillatory solution of this type has been obtained for GR case in Ref. \cite{Ellisbook1}. This behavior is also observed in Fig. \ref{Figtwofluidradiation123}.
We therefore say that for $n>1$, we have oscillator grow in amplitude
radiation fluctuations. But if one looks at the case of $n=0.31$, that is $0<n<1$, (see Fig. \ref{Figtwofluidradiation031} ) 
we realized that the perturbations have high amplitude and later
decays in oscillatory way. Therefore, we can say that the weaker gravity $(n<1)$, tend not to have growing radiation perturbations, yet the 
high gravity, that is $n>1$, clearly has a hand on the growth of the radiation perturbations.

\section{Conclusions and discussions} \label{Conclusion1}
In this paper we have studied linear cosmological perturbations of a flat FLRW spacetime background using 
the $1+3$ covariant and gauge-invariant formalism, where the Brans-Dicke scalar-tensor theory is the underlying model
of gravitation assumed.  We have used the equivalence between the scalar-tensor and $f(R)$ gravity theories to develop perturbation
equations based on gradient variables defined in a covariant and gauge-invariant way. The extra degree of freedom that differentiates $f(R)$ 
from GR is defined to be  an extra scalar-field degree of freedom.  Thus the density perturbations considered correspond to a scalar 
field--standard matter two-fluid cosmic medium.  We have then specialized to a scalar field-dust and scalar field-radiation fluid systems 
to analyse some interesting solutions. 
The energy density perturbations of dust grow exponentially in time for different values of $n$. This result is in agreement with the existing literature about perturbations of $R^{n}$ gravity toy models.
For the case of the radiation-dominated epoch, the behavior of the amplitude of the radiation perturbations  depends on the nature of $n$. 
For $n$ less than unity, the perturbations decay with time in oscillatory mode, but for the 
values of $n$ greater than unity, the fluctuations grow and oscillate. This radiation perturbations behavior is studied 
under long-wavelength assumption.
A more detailed analysis is needed in the context of non-vanishing scalar field perturbations.

\section*{Acknowledgments} 
JN gratefully acknowledges financial support from the Swedish International Development Cooperation Agency (SIDA) 
through the International Science Program (ISP) to the University of Rwanda (Rwanda Astrophysics, Space and Climate Science Research Group),
and Physics Department, North-West University, Mafikeng
Campus, South Africa, for hosting him during the preparation of this paper.
AA  acknowledges that this work is based on the research supported in part by the National Research Foundation of South Africa and the Faculty Research Committee of the Faculty of Natural and Agricultural Sciences of North-West University.

\bibliographystyle{unsrt}

\section*{Appendix}
In this Appendix, we present the relationship between curvature gradient $\mathcal{R}_{a}$ and $\Phi_{a}$ as
\begin{equation}
 \mathcal{R}_{a}=\frac{\Phi_{a}}{\phi'}\; .\label{relationRaandPhia}
\end{equation}
The relationship between curvature momentum gradient $\Re_{a}$ and, $\Phi_{a}$ and scalar field momentum gradient $\Psi_{a}$ is given
\begin{equation}
\Re_{a}=\frac{\Psi_{a}}{\phi'}-\frac{\dot{\phi}\phi''}{\phi'^{3}}\Phi_{a}\; ,\label{relationReandPsia}
\end{equation}
where $\mathcal{R}_{a}=a\tilde{\nabla}_{a}R$ and $\Re_{a}=a\tilde{\nabla}_{a}\dot{R}$.
We provide equations relating covariant quantities between both theories: $f(R)$ and Scalar tensor theories.
We start by obtaining
the scalar variables from equations \eqref{relationRaandPhia} and \eqref{relationReandPsia} as
\begin{equation}
 \mathcal{R}=\frac{\Phi}{\phi'}\; ,\label{relationRaandPhia1}
\end{equation}
\begin{equation}
\Re=\frac{\Psi}{\phi'}-\frac{\dot{\phi}\phi''}{\phi'^{3}}\Phi\; ,\label{relationReandPsia1}
\end{equation}
where $\mathcal{R}=a\tilde{\nabla}^{a}\mathcal{R}_{a}$ and $\Re=a\tilde{\nabla}^{a}\Re_{a}$.
Thus one has the evolution equations of perturbation quantities as
\begin{equation}
\dot{\Delta}_{m}=-(1+w)Z+w\Theta \Delta_{m}\; , \label{f(R)dotDeltasingle1}
\end{equation}
\begin{equation}
\begin{split}
\dot{Z}&=\Big(\frac{\dot{\phi}}{\phi+1}-\frac{2\Theta}{3}\Big)Z-\frac{w}{(1+w)}\tilde{\nabla}^{2}\Delta_{m}
-\frac{\phi'}{\phi+1}\tilde{\nabla}^{2}\mathcal{R}+\Big[\frac{1}{2\phi'}+\frac{\mu_{m}}{(\phi+1)^{2}}-\frac{f}{2(\phi+1)^{2}}\\
&-\frac{\Theta\dot{\phi}}{(\phi+1)^{2}}-\frac{\phi''\tilde{\nabla}^{2}R}{\phi'(\phi+1)}
+\frac{\phi'\tilde{\nabla}^{2}R}{(\phi+1)^{2}}-\frac{2K}{(\phi+1)a^{2}}
+\frac{\Theta\dot{\phi}\phi''}{\phi'^{2}(\phi+1)}\Big]\phi'\mathcal{R}+\frac{\Theta\phi'}{\phi+1}\Re\\
&+\Big[\frac{\phi w-1}{(w+1)(\phi+1)}\mu_{m}-\frac{w}{1+w}\Big(-\frac{1}{3}\Theta^{2}
+\frac{f}{2(\phi+1)}+\frac{\Theta\dot{\phi}}{\phi+1}
-\frac{\phi'}{\phi+1}\tilde{\nabla}^{2}R\Big)\Big]\Delta_{m}\; ,\label{f(R)dotZsingle1}
\end{split}
\end{equation}
%%%%%%%%%%%%%%%%
\begin{equation}
\dot{\mathcal{R}}=-\Big(\frac{\dot{\phi}'}{\phi'}-\frac{\dot{\phi}\phi''}{\phi'}\Big)\mathcal{R}+\phi'\Re
-\frac{w\dot{\phi}}{w+1}\Delta_{m}\; ,\label{f(R)dotPhisingle1}
\end{equation}
\begin{equation}
\begin{split}
 \dot{\Re}&=-\Big\{\frac{\dot{\phi}'}{\phi'}+\frac{\dot{\phi}\phi''}{\phi'}
 -\frac{\dot{\phi}\phi''}{\phi'}\Big(\dot{\phi}'+\frac{\dot{\phi}'}{\phi'}-2\Big)-\frac{\dot{\phi}'}{\phi'}
 \Big\}\Re-\Big\{\frac{\ddot{\phi}\phi''}{\phi'^{2}}+\frac{\dot{\phi}\dot{\phi}''}{\phi'^{2}}
 -\frac{\dot{\phi}\dot{\phi}'\phi''}{\phi'^{3}}-\frac{\dot{\phi}\phi''}{\phi'^{2}}\Big(\frac{\dot{\phi}'}{\phi'}\\
& -\frac{\dot{\phi}\phi''}{\phi'}\Big) -\Big[\frac{\dot{\phi}\phi''}{\phi'}\Big(\dot{\phi}'+\frac{\dot{\phi}'}{\phi'}-2\Big)
 +\frac{\dot{\phi}'}{\phi'}\Big]\frac{\dot{\phi}\phi''}{\phi'}+\frac{\phi''}{\phi'}\Big[\frac{\dot{\phi}^{2}}{\phi'}\Big(1
 +\frac{1}{\phi'}\Big)(\frac{\phi'''}{\phi''}-\frac{\phi''}{\phi'})
 -\frac{\dot{\phi}^{2}}{\phi'}\frac{\phi'''}{\phi''}\\
 &+\frac{\dot{\phi}^{2}\phi''}{\phi'^{2}} -\frac{\dot{\phi}\dot{\phi}'}{\phi'^{2}}\Big] \Big\}\mathcal{R}
+\Big(\frac{w\dot{\phi}^{2}\phi''}{\phi'^{2}(w+1)}-\frac{w\ddot{\phi}}{\phi'(w+1)}\Big)\Delta_{m}\; ,\label{f(R)dotPsisingle1}
\end{split}
\end{equation}
and constraint equation is given as
\begin{equation}
\begin{split}
\frac{C}{a^{2}}&=-\Big(\frac{4}{3}\Theta +\frac{2\ddot{\phi}}{\phi+1}\Big)Z+\frac{2\mu_{m}}{\phi+1}\Delta_{m}
+\frac{2\phi'}{\phi+1}\tilde{\nabla}^{2}\mathcal{R}+\Big\{\Big[-\frac{2\mu_{m}}{(\phi+1)^{2}}+\frac{f}{(\phi+1)^{2}}\\
&-\frac{2\Theta}{\phi+1}\frac{\phi''}{\phi'}\Big(\frac{\dot{\phi}^{2}}{\phi'}(\frac{\phi'+1}{\phi'})(\frac{\phi'''}{\phi''}-\frac{\phi''}{\phi'})
-\frac{\dot{\phi}^{2}}{\phi'}\frac{\phi'''}{\phi''}
 \frac{\dot{\phi}^{2}\phi''}{\phi'^{2}}-\frac{\dot{\phi}\dot{\phi}'}{\phi'^{2}}\Big)
 +\frac{2\Theta\ddot{\phi}}{\phi+1}+\frac{4K}{a^{2}(\phi+1)}\\
& +\frac{2\phi''\tilde{\nabla}^{2}R}{(\phi+1)\phi'}-\frac{2\phi'\tilde{\nabla}^{2}R}{(\phi+1)^{2}}\Big]\phi'
-\frac{2\Theta}{\phi+1}\Big[\Big(\dot{\phi}+\frac{\dot{\phi}}{\phi'}\Big)\frac{\phi''}{\phi'}
 +\dot{\phi}'-\frac{2\dot{\phi}\phi''}{\phi'}+\frac{\dot{\phi}'}{\phi'}\Big]\frac{\dot{\phi}\phi''}{\phi'}\Big\}\mathcal{R}\\
&-\frac{2\Theta}{\phi+1}\Big[\Big(\dot{\phi}+\frac{\dot{\phi}}{\phi'}\Big)\frac{\phi''}{\phi'}
 +\dot{\phi}'-\frac{2\dot{\phi}\phi''}{\phi'}+\frac{\dot{\phi}'}{\phi'}\Big]\phi'\Re \; .\label{Csingle1f(R)}
\end{split}
\end{equation}

%%%%%%%%%%%%%%%%%%%%%%%%%%%%%%%%%%%%%%%%%%
%=====================================
% \bibliography{./references}
% %\renewcommand{\bibname}{References}
% %\nocite{*}%\bibliographystyle{revcompchem}
% %\bibliographystyle{naturemag}
% %\bibliographystyle{amsalpha} %% acm, naturemag, revcompchem
% %\bibliographystyle{alpha}
% %\bibliographystyle{unsrt}
% %\bibliographystyle{apalike}
% %\bibliographystyle{amsplain}
% %\bibliographystyle{plain}
% %\bibliographystyle{ieeetr}
% %\bibliographystyle{h-physrev3.bst}
% %\bibliographystyle{amsplain}
% %\bibliographystyle{abbrv}
% %\bibliographystyle{apsrev}
% \bibliographystyle{iopart-num}
\end{document}